\documentclass[seceq]{ptptex}

\usepackage{graphicx}

\def\SU{\mathop{\rm SU}}
\def\SO{\mathop{\rm SO}}
\def\SL{\mathop{\rm SL}}
\def\U{\mathop{\rm U}}

\title{
Baryon vertices in AdS black hole backgrounds%
}

\author{
Yosuke \textsc{Imamura}\thanks{E-mail: \tt imamura@hep-th.phys.s.u-tokyo.ac.jp}%
}

\inst{
Department of Physics, University of Tokyo, Tokyo 113-0033, Japan
}

\abst{
We numerically computed the energy of QCD string junctions
in non-supersymmetric $\SU(N)$ gauge theories
as the energy of brane configurations in
the background of AdS black hole solutions,
and extracted the contribution of baryon vertices.
We obtain a negative vertex energy for the three- and four-dimensional Yang-Mills
theories and a vanishing vertex energy for the five-dimensional theory.
}

\begin{document}

\maketitle

\section{Introduction}
Although string theory was originally constructed for the purpose
of describing the behavior of quarks in hadrons,
it has been developed for long time independently of
hadron physics
as a possible form for a theory unifying
gravity and other interactions,
because it can be formulated only in $10$ or $26$
spacetime dimensions,
and spin $2$ massless particles arise naturally.
The scale of the fundamental string tension, $1/(2\pi\alpha')$,
which is near the Planck scale,
was thought to have no relation to
the QCD scale $1/\alpha'_{\rm QCD}\sim{\rm GeV}^2$.

The discovery of the AdS/CFT correspondence\cite{maldacena,GKP,holography}
changed this situation.
In this correspondence, string theory in a $d+1$-dimensional AdS or AdS-like spacetime,
which is usually accompanied by an internal space,
is dual to a gauge theory
in the $d$-dimensional boundary of the spacetime.
This enables us to study the non-perturbative dynamics
of gauge theories with the help of string theory.
The properties of the gauge theory are reflected by the
structure of the $d+1$-dimensional spacetime and the internal space.
The dual geometry of the confining gauge theory is a kind of (Wick-rotated)
black hole, and
the QCD strings in the boundary theory are identified with fundamental
strings in the region near the horizon of the black hole.
(The term ``horizon'' here is not used in its ordinary sense,
because of the Wick rotation.
On the ``horizon'', some space-like components of the metric
vanish, while the time component $g_{tt}$ is regular.)
The difference between
the QCD string tension, $1/(2\pi\alpha'_{\rm QCD})$, and
the fundamental string tension, $1/(2\pi\alpha')$,
is due to a large red-shift associated with the curved background.

In this paper, we consider large $N$ $\SU(N)$ pure Yang-Mills theory in $d=3$ and $4$ dimensions.
The AdS dual of this theory is the $d+1$-dimensional AdS Schwarzchild black hole solution\cite{thermal}.
We also consider the $d=5$ case, for reasons
discussed in \S\ref{num.sec}.
The $5$-dimensional gauge theory is
non-renormalizable, and we define it as the low-energy effective theory
of open strings on D5-branes wrapped on non-supersymmetric ${\bf S}^1$.

Let $g[C]$ denote the path-ordered integral
defined for a general open contour $C$ by
\begin{equation}
g[C]={\rm P}\exp\int_CA\in\SU(N).
\label{gcdef}
\end{equation}
If the two ends of $C$ coincide,
the Wilson loop associated with the closed loop $C$
and the $\SU(N)$ representation $R$
is defined by $W_R[C]\equiv \langle\rho_R(g[C])^a{}_a\rangle$,
where $\rho_R(g)^a{}_b$ is the representation matrix of $g\in\SU(N)$
in the representation $R$.
Let us take a rectangle of width $L$ along the spatial direction
and length $t$ along the temporal direction as the contour $C$.
If the charge associated with the representation $R$ is confined,
the free energy $\Gamma$ defined
by $W_R=e^{-t\Gamma}$ for large $t$ takes the form
\begin{equation}
\Gamma=LT+2a+{\cal O}(L^{-1}).
\label{Emeson}
\end{equation}
This can be regarded as the energy of a
``meson'' consisting of the charges $R$ and $\overline R$
connected by a QCD string.
The quantity $T$ is interpreted as the tension of the QCD string,
and $a$ represents the boundary correction
for each end of the string.
The representation $R$ is conveniently expressed in the form of  by
a Young tableau.
Due to the vacuum polarization of
the gluon field,
information concerning the shape of this Young tableau is screened,
and only the mod $N$ part of the number of boxes in the tableau
can be observed from a distance.
This number is called the $N$-ality of the representation $R$.
We refer to strings associated with a representation of
an $N$-ality $k$ as ``$k$-strings.''

Let $T(k)$ denote the tension of a $k$-string.
It has a non-linear dependence on $k$ non-linearly, and it satisfies
the relations
\begin{equation}
T(0)=0,\quad
T(k+nN)=T(k),\quad
T(k)=T(-k).
\end{equation}
The last relation here follows from the fact that
($-k$)-strings are $k$-strings with the opposite orientation.
Let us assume that the function $T(k)$ is convex except at $k=0$ ($\mod N$).
This is natural because $T(0)=T(N)=0$ and $T(k)>0$ for $0<k<N$.
The convexity assumption implies the inequality
\begin{equation}
T(k_1)+T(k_2)>T(k_1+k_2).\quad
(k_1, k_2\neq0\mod N)
\end{equation}
This means that two QCD strings form a truly bound state.
Thus we can regard
a $k$-string as a bound state of
$k$ $1$-strings.

Through the AdS/CFT correspondence, a Wilson loop with $k$-ality $\pm 1$
is mapped to
a fundamental string with boundary $C$, which is embedded in the boundary
of the AdS space.
For the purpose of obtaining the QCD string tension with the help of
AdS/CFT, we consider a long string
most of which lies on the horizon.
The QCD string tension is the fundamental string tension multiplied by
the red-shift factor at the horizon.
This analysis was first performed for the non-confining ${\cal N}=4$ Yang-Mills theory in Ref. \citen{wl},
and it was confirmed that there is no linear term.
Such Wilson lines were also studied in Ref. \citen{reyyee} in
a different way.
The analysis in confining theories was first carried out in Ref \citen{thermal},
and a non-vanishing QCD string tension was
obtained.
Only the horizon in the spacetime is
important when we compute the QCD string tension.
We call the horizon ``IR subspace.''

For a large class of solutions dual to confining field theories,
the IR subspace has the common structure $M_d\times {\bf S}^p$,
where $M_d$ is the $d$-dimensional flat Minkowski space
and ${\bf S}^p$ is a $p$-dimensional sphere.
The internal sphere ${\bf S}^p$, through which flux passes,
plays an important role when we consider objects dual to $k$-strings
with general $k$.
In Ref. \citen{tension} it was shown that
$k$ coincident fundamental strings are expanded into a D$p$-brane tube due to
the Myers effect\cite{myers},
and the tension of $k$-strings is equal to the energy of the D$p$-brane tube
per unit length.

The meson configuration discussed above
can easily be generalized to general hadron configurations.
For example, we can construct a baryon configuration consisting of
three charges $R_i$ ($i=1,2,3$) in the following way.
Let $C_i$ ($i=1,2,3$) be three open contours sharing the same endpoints.
Here we assume that the endpoints are at $(t,\vec r)=(0,\vec 0)$ and $(t,\vec 0)$,
and that each $C_i$ consists of three segments,
$(0,\vec 0)$-$(0,\vec r_i)$,
$(0,\vec r_i)$-$(t,\vec r_i)$ and
$(t,\vec r_i)$-$(t,\vec 0)$.
We define the $\SU(N)$-valued operator $g[C_i]$ by
(\ref{gcdef}) for each $C_i$, and we combine them to form the gauge singlet observable
\begin{equation}
B=\langle T_{abc}T^{*a'b'c'}
\rho_{R_1}(g[C_1])^a{}_{a'}
\rho_{R_2}(g[C_2])^b{}_{b'}
\rho_{R_3}(g[C_3])^c{}_{c'}
\rangle,
\end{equation}
where $T_{abc}$ is an $\SU(N)$ invariant tensor
whose three indices correspond to the representations $R_1$, $R_2$ and $R_3$,
and $T^{*a'b'c'}$ is its complex conjugate.
If all the charges $R_i$ are confined,
three QCD strings associated with these charges arise.
They are attached to the contours $C_i$ at one end,
and the other ends of the three strings are joined at a so-called baryon vertex.
These three strings possess the string charges $k_i$, which are the
$N$-alities of the representations $R_i$.
The existence of the invariant tensor $T_{abc}$ requires that $k_i$
satisfy the condition
\begin{equation}
k_1+k_2+k_3=0.\quad(\mod N)
\end{equation}
This can be interpreted as the mod $N$ conservation law of
the string charge.
We conjecture that the free energy of this baryon configuration
defined by $B=e^{-t\Gamma}$
takes the form
\begin{equation}
\Gamma=\sum_{i=1}^3(L_iT(k_i)+a(R_i))+E_{\rm vertex}(k_1,k_2,k_3)+{\cal O}(L_i^{-1}),
\label{gammaB}
\end{equation}
where $T(k_i)$ is the string tension of the $i$-th string, and
$a(R_i)$ is the boundary correction associated with the contour $C_i$.
These two kinds of quantities are the same as those in the meson energy (\ref{Emeson}).
The quantity $L_i$ is the length of the $i$-th string.
Note that this length may be different from $|\vec r_i|$,
because the baryon vertex is in general not at the origin.
The position of the vertex is kinematically determined by the balance of the tensions of
the three strings.

The $E_{\rm vertex}$
in (\ref{gammaB}) denotes the contribution of the baryon vertex,
and it is a new ingredient, which is absent in the meson case.
On the gravity side, the object corresponding to a baryon vertex is
a D$p$-brane wrapped on the internal ${\bf S}^p$\cite{baryon,ooguri}.
Because both QCD strings and baryon vertices are realized
with D$p$-branes, the entire system of a junction
is dual to a smooth D$p$-brane configuration.
Thus, it is possible to compute the energy of
junctions and extract $E_{\rm vertex}$ from this energy by using
the Born-Infeld action and the Chern-Simons action of the D$p$-brane.
In a previous work of the present author\cite{junc},
$E_{\rm vertex}$ was computed for the Maldacena-N\'u\~nez
and Klebanov-Strassler backgrounds,
both of which are dual to ${\cal N}=1$
large $N$ confining gauge theories.
In this paper, we report the result of the same analysis for different backgrounds
without supersymmetry.

\section{AdS black hole solution}\label{black.sec}
In order to realize the $d$-dimensional pure $\SU(N)$ Yang-Mills theory,
we consider $N$ coincident D$d$-branes wrapped
on a non-supersymmetric ${\bf S}^1$ cycle\cite{thermal}.
A classical solution in Type II supergravity
corresponding to this brane configuration
is obtained by applying double Wick rotation to the non-extremal D$d$-brane solution
constructed in Ref. \citen{HS},
and its near horizon geometry is the AdS Schwarzchild black hole.
The metric in the string frame is
\begin{equation}
ds^2=
      f_-^{\frac{1}{2}}\eta_{ij}dX^idX^j
     +f_+f_-^{-\frac{1}{2}}dy^2
     +f_+^{-1}f_-^{\frac{7-3p}{2(p-1)}}dr^2
     +f_-^{\frac{5-p}{2(p-1)}}r^2d\Omega_p^2,
\label{Dpmetric}
\end{equation}
where $p=8-d$, and $f_\pm\equiv f_\pm(r)$ are the following functions of $r$:
\begin{equation}
f_\pm(r)=1-\frac{r_\pm^{p-1}}{r^{p-1}}.
\end{equation}
The two parameters $r_-$ and $r_+$ represent the locations of
the inner and outer horizon, respectively, and they satisfy
the relation
\begin{equation}
\sqrt{r_+^{p-1}r_-^{p-1}}
=\frac{Ng_{\rm str}^\infty(2\pi l_s)^{p-1}}{(p-1)\Omega_p},
\end{equation}
where we define $l_s$ by $l_s\equiv\alpha'^{1/2}$,
and $g_{\rm str}^\infty$ denotes the string coupling constant
in the asymptotic region, $r\rightarrow\infty$.
The coordinate $y$ is periodic, and its period is determined by
requiring the solution to be smooth at the horizon, $r=r_+$.
Let the period be $2\pi R_y$.
Then the Kaluza-Klein radius $R_y$ is given by
\begin{equation}
R_y=\frac{2}{p-1}\frac{r_+^{(p-1)/2}}{\Delta^{(p-3)/2}},
\label{rdr2}
\end{equation}
where $\Delta$ is defined by $\Delta^{p-1}\equiv r_+^{p-1}-r_-^{p-1}$.
The $r$ dependence of the dilaton field in the classical solution is given by
\begin{equation}
e^\phi=g_{\rm str}^\infty f_-^{\frac{5-p}{4}}(r).
\end{equation}
There exists the non-vanishing RR flux in ${\bf S}^p$
\begin{equation}
G_p=\frac{N}{\Omega_p}\omega_p,
\label{bkgflux}
\end{equation}
where $\Omega_p$ is the volume of the unit ${\bf S}^p$, and
$\omega_p$ is the volume form on ${\bf S}^p$ normalized
such that $\oint\omega_p=\Omega_p$.

QCD strings in the boundary gauge theory are identified with
fundamental strings
or D-branes possessing the fundamental string charge
in this spacetime.
Because of the gravitational force,
they fall to the bottom, $r=r_+$, in this spacetime.
Thus, only the subspace $r=r_+$, which we call the IR subspace,
is relevant to our analysis.
The metric of the IR subspace is
\begin{equation}
ds^2
=-dt^2+R^2((dx^a)^2+d\Omega_p^2),\quad
R=f_-^{\frac{5-p}{4(p-1)}}(r_+)r_+,
\label{Rdefin}
\end{equation}
where we rescale the coordinates $X^i$
in (\ref{Dpmetric})
and denote the temporal and spatial coordinates by $t$ and $x^a$ ($a=1,\ldots,p$), respectively.

In order to decouple gravity from the boundary gauge theory, we need to take the decoupling limit, $l_s\rightarrow 0$, keeping the parameters in the gauge theory fixed.
If we require the QCD string tension to remain finite in this limit,
the redshift factor $f_-^{1/2}(r_+)$ in the metric (\ref{Dpmetric}),
which relates the fundamental string tension $\sim1/\alpha'$ and
the QCD string tension $\sim1/\alpha'_{\rm QCD}$, should behave as
\begin{equation}
(g_{tt}(r_+))^2=f_-(r_+)=1-\frac{r_-^{p-1}}{r_+^{p-1}}\sim\frac{\alpha'^2}{\alpha'^2_{\rm QCD}}
\rightarrow0.
\label{decoupling}
\end{equation}

\section{QCD strings from AdS/CFT}
The mod $N$ nature and the formation of bound strings
can be understood on the AdS side by taking account of the
transition between fundamental strings
and D$p$-branes in the background RR flux
(\ref{bkgflux}).
In the IR subspace, fundamental strings are blown up to
D$p$-brane tubes with electric flux on their world volumes by the
Myers effect\cite{myers}.
The energy of a D$p$-brane is given by
\begin{equation}
E
 =T_{{\rm D}p}R^p\int d^p\sigma
 \sqrt{\det g_{ab}}\sqrt{1+ g_{ab}D^aD^b},
\label{Denergy}
\end{equation}
where $g_{ab}$ and $D^a$ ($a,b=1,\ldots,p$) are
the induced metric
and the electric flux density, respectively, on the D$p$-brane worldvolume.
We factor out the radius $R$ from the metric to simplify the equations
in what follows.
The normalization of $D^a$ is chosen so that the expression for the D-brane
energy is simplified as (\ref{Denergy}).
The energy (\ref{Denergy}) does not include parameters for the background geometry, except for an
overall factor that does not affect the
equations of motion of the D$p$-branes.

A D$p$-brane tube has a cross section ${\bf S}^{p-1}$ wrapped on
a contractible cycle in ${\bf S}^p$.
The behavior of this cross section
is analogous to that of a flexible superconducting ring in a magnetic
flux background.
In general, electric current exists on the ring.
In this analogy, the role of the background magnetic field is played by
the RR flux, and the role of the electric current on the ring
by the electric flux $D^a$ on the D$p$-brane tube.
If the cross section of the D-brane tube (which is topologically ${\bf S}^{p-1}$
in the internal space ${\bf S}^p$)
changes,
the electric flux $D^a$ changes
due to an effect analogous to the electromagnetic induction.
In this process, the quantity
\begin{equation}
\frac{k}{N}
 =\frac{b}{(p-1)\Omega_p}
   \oint_{{\bf S}^{p-1}}D^adS_a
 +\frac{1}{\Omega_p}\int_{{\bf D}^p}\omega_p
\label{wtn}
\end{equation}
is invariant, where the parameter $b$ is defined by
\begin{equation}
b=(p-1)\Omega_p\frac{T_{Dp}R^{p-1}}{NT_{\rm str}}.
\label{bdef}
\end{equation}
The first term in (\ref{wtn})
gives the total electric flux passing through the
cross section of the D-brane tube and the second term is the volume enclosed
by the cross section.
The quantity $k$ defined by (\ref{wtn}) is identified with the string charge.
Indeed, if the tube shrinks down to a string and the second term
in (\ref{wtn}) vanishes, $k$ is simply the amount of the electric flux
that couples to the NS-NS $2$-form field.
The mod $N$ nature is reflected in
the ambiguity in the
definition of the inside of the cross section, due to
the nontrivial homotopy $\pi_p({\bf S}^p)={\bf Z}$.

The equation (\ref{wtn}) not only defines
the charge $k$ but also places a constraint
on the electric flux.
The differential form of (\ref{wtn})
is $(b/(p-1))d*D+\omega_p=0$,
and this is the Gauss constraint imposed on $D^a$.
Stable brane configurations are determined by minimizing the energy
(\ref{Denergy})
under this constraint.

The parameter $b$ is the only parameter of the background
spacetime relevant to our analysis.
Roughly speaking, this represents the ``toughness'' of D$p$-branes
in the following sense.
To analyze a baryon configuration,
we need to consider a baryon vertex joining strings,
as mentioned in the introduction.
The vertex is a D$p$-brane wrapped on the ${\bf S}^p$\cite{baryon,ooguri}.
It is deformed by the tension of the strings attached to it.
If the number of strings is of order $N$,
the tension pulling on the D$p$-brane is of the order of $NT_{\rm str}$.
The scale of the deformation $L$ is determined by the ratio between this tension
and the D-brane tension $T_{{\rm D}p}$ as
$L^{-(p-1)}=T_{Dp}/(NT_{\rm str})$.
If we normalize this with the size of the ${\bf S}^p$,
we obtain the dimensionless parameter $b$ defined above.
If $b$ is very large, the D$p$-brane is ``tough'', and
the deformation is negligible compared to the
size of the baryon vertices, while if $b$ is of order $1$ or larger,
the brane is ``soft'', and the deformation caused by the string tension should be taken into account.

Let us determine the value of the parameter $b$ for
the IR subspace of the AdS black hole solution
in \S\ref{black.sec}.
The fundamental string tension and the D$p$-brane tension
in the IR subspace are
\begin{equation}
T_{\rm str}=\frac{1}{2\pi l_s^2},\quad
T_{{\rm D}p}=\frac{1}{(2\pi)^pl_s^{p+1}g^{\rm hor}_{\rm str}},
\end{equation}
where $g_{\rm str}^{\rm hor}=g_{\rm str}^\infty f_-^{\frac{5-p}{4}}(r_+)$ is the string coupling constant
at $r=r_+$.
Substituting these and $R$ in (\ref{Rdefin}) into
the definition (\ref{bdef}) of the parameter $b$,
we obtain
\begin{equation}
b
=\left(\frac{r_+}{r_-}\right)^{\frac{p-1}{2}}.
\end{equation}
In the decoupling limit (\ref{decoupling}),
this becomes $1$ for any value of $p$.
Thus we have to take account of the deformation of D-branes.

\begin{figure}[htb]
\centerline{\includegraphics[width=.6\linewidth]{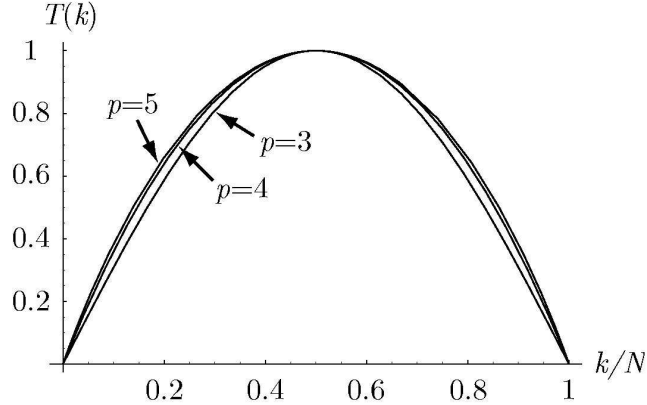}}
\caption{The $k$ dependence of the string tension for $p=3,4,5$.
We normalize the tension such that $T(N/2)=1$.}
\label{tplot.eps}
\end{figure}

Now we are ready to compute the QCD string tensions
for $p=3,4,5$.
Let $\theta$ be the angular radius of the cross section of the tube.
For given $k$, the energy of the D$p$-brane tube is minimized at the value of $\theta$ satisfying
\begin{equation}
\frac{k}{N}
=\frac{\Omega_{p-1}}{\Omega_p}\left[\frac{b^2}{p-1}\cos\theta\sin^{p-2}\theta
+\int_0^\theta\sin^{p-1}xdx\right],
\label{minimumT}
\end{equation}
and the tension is given by
\begin{equation}
T(k)
=T_{{\rm D}p}\Omega_{p-1}R^{p-1}\sin^{p-2}\theta
\sqrt{\sin^2\theta+b^2\cos^2\theta}
=T_{{\rm D}p}\Omega_{p-1}R^{p-1}\sin^{p-2}\theta,
\end{equation}
where we have used $b=1$ in the final step.
The tension
for $p=3$, $4$, $5$, 
is given as follows:
\begin{eqnarray}
T_{p=3}(k)&=&T_{{\rm D}3}\Omega_2R^2\sin\left(\frac{\pi k}{N}\right),
\label{p3tension}\\
T_{p=4}(k)&=&T_{{\rm D}4}\Omega_3R^3\frac{4k(N-k)}{N^2},
\label{p4tension}\\
T_{p=5}(k)&=&T_{{\rm D}5}\Omega_4R^4\sin^3\theta
\nonumber\\&&\quad\mbox{with $\theta$ solving}
\quad
\pi(k/N)=\theta-\cos\theta\sin\theta.
\end{eqnarray}
(In order to obtain finite QCD string tensions in the decoupling limit,
we need to rescale these tensions by appropriate warp factors.
We here do not include them, because
we are interested in the relative magnitude of the vertex energies
and the energies of the corresponding wrapped branes.)
A plot of the function $T(k)$ is given in Fig. \ref{tplot.eps}.
As seen there,
the functions $T(k)$ for different $p$ are quite similar.

\section{Numerical analysis of junctions}\label{num.sec}
In this section, we report the results of numerical computation of the energy of baryon vertices
for $p=3,4,5$ carried out with the following steps.
\begin{itemize}
\item
Firstly, we determine
the directions $\phi_i$ ($i=1,2,3$)
of the three strings in a junction
by imposing the condition
that the three string tensions balance.
If the charges of the three strings are $k_i$,
the angles $\phi_i$ are determined by
\begin{equation}
\cos(\phi_2-\phi_3)
=\frac{T^2(k_2)+T^2(k_3)-T^2(k_1)}{2T(k_2)T(k_3)},
\end{equation}
and similar equations obtained by permuting the subscripts
$1$, $2$ and $3$.
\item
Secondly, we construct an initial configuration
consisting of three strings for our numerical computation.
The three strings are joined at one end of each string by
a baryon vertex, and the other ends
are pinned
on a circle with a sufficiently large radius $L$.
We use $L=4R$ in the analysis below.
The positions on the circle are chosen as
$(L\cos\phi_i,L\sin\phi_i)$ ($i=1,2,3$),
using the angles obtained in the first step,
so that the baryon vertex coincides with the center of the circle.
\item
Finally, we determine the brane configurations
minimizing the energy.
Then we extract the energy $E_{\rm vertex}$ of vertex
from the energy $E_{\rm junc}$ of the junction
using the relation
\begin{equation}
E_{\rm vertex}=E_{\rm junc}-\sum_{i=1}^3LT(k_i).
\label{extract}
\end{equation}
\end{itemize}

As preparation for the numerical analysis,
we should reduce the problem to a lower dimensional one.
We consider only planar junctions in the $(x,y)$-plane.
We embed the internal space, ${\bf S}^p,$ into ${\bf R}^{p+1}$ using the Cartesian coordinates $(u,v,w_1,\ldots,w_{p-1})$.
Then, we assume that the D$p$-brane worldvolumes,
which are branched tubes with cross section ${\bf S}^{p-1}$,
are invariant under the $\SO(p-1)$ rotation in
the $(w_1,\ldots,w_{p-1})$-space,
and we represent a worldvolume as an $\SO(p-1)$ orbit of
a two-dimensional surface in the $(x,y,u,v)$-space.
Next, we introduce the flux density $D'^a$ on the surface
and the surface integral measure $dn_a$ by
\begin{equation}
D'^a=w^{p-2} D^a\quad (a=u,v),\quad
dS_a=\Omega_{p-2}w^{p-2}dn_a,
\end{equation}
where $w=\sqrt{\sum_iw_i^2}=\sqrt{1-u^2-v^2}$.
The energy of a brane and the integral form of the Gauss law constraint for the electric flux are given by
\begin{equation}
\frac{E}{T_{{\rm D}p}R^p}
=\Omega_{p-2}\int d^2\sigma\sqrt{\det g_{ab}}
               \sqrt{w^{2(p-2)}+ g_{ab}D'^aD'^b},
\label{H2dim}
\end{equation}
\begin{equation}
2\pi\frac{k}{N}
 =b\int dn_aD'^a
 +(p-1)\int w^{p-3}du\wedge dv.
\label{wtnconst}
\end{equation}
The problem we consider is to find two-dimensional surfaces
that minimize the energy (\ref{H2dim}) under the constraint (\ref{wtnconst}),
and then to compute
these minimal energies.
We realize the two-dimensional surface as a mesh
consisting of triangles.
We give only the result below.
(See Ref.\citen{junc} for the details of the numerical analysis.)

\subsection{$p=3$}\label{MNKS.sec}
The classical solution with $p=3$ describes the near-horizon geometry
of coincident D5-branes wrapped around non-supersymmetric ${\bf S}^1$.
Because the five-dimensional gauge theory is non-renormalizable,
it should be regarded as the low-energy effective theory
of open strings on the D5-branes.

There are other reasons that the $p=3$ case is interesting.
By performing the S-duality transformation, we obtain the NS5-brane
solution, and its near horizon geometry is
\begin{equation}
ds^2=\eta_{ij}dX^idX^j
+r_+^2\left[h(\rho)d\theta^2+h^{-1}(\rho)\frac{d\rho^2}{\rho^2}+d\Omega_3^2\right],\quad
g_{\rm str}=g_{\rm str}^\infty\frac{r_+}{\rho},
\end{equation}
where $h(\rho)=1-\Delta^2/\rho^2$,
and we have introduced the angular variable $\theta=y/r_+$ with period $2\pi$.
It is known that string theory in this background is solvable.
The two-dimensional space parameterized by $(\theta,\rho)$ constitutes a cigar geometry
and is described by the supersymmetric
level $N$ $\SL(2,{\bf R})/\U(1)$ CFT with central charge $c=3(N+2)/N$.
The ${\bf S}^3$ part can be described by
the supersymmetric $\SU(2)_N$ WZW model with the central charge $c=3(N-2)/N+3/2$.

Another reason to study the $p=3$ case is that
the vertex energy in this case can be computed analytically.
The IR subspace with $p=3$ and $b=1$
is also regarded as the IR subspace of the Maldacena-N\'u\~nez solution\cite{MN1,MN2},
which is the near horizon geometry of the D5-branes wrapped around the $2$-cycle
of a conifold.
The Maldacena-N\'u\~nez solution is believed to be dual to the ${\cal N}=1$ Yang-Mills theory.
In Ref. \citen{junc}, it was analytically shown that
the energies of vertices in this case vanish.
Although we know the exact result,
we present the result of the numerical analysis
in order to clarify the accuracy of the method.
We use $R^3T_{{\rm D}3}$ as the unit of energy.
\begin{table}[htb]
\caption{The energies of baryon vertices for $p=3$ and $b=1$.
The first column lists the ratio of charges of the three strings in a junction.
The next three columns list the energies of the junctions realized as triangular meshes
consisting of
$n_{\rm mesh}=6\times20^2$,
$n_{\rm mesh}=6\times40^2$ and
$n_{\rm mesh}=6\times80^2$ triangles, respectively.
The energy $E_\infty$ is obtained by extrapolating these results to the continuum limit $n_{\rm mesh}\rightarrow\infty$.
The quantity $E_0$ represents the energy of strings defined by $E_0=L\sum_{i=1}^3T(k_i)$.
The last column shows the energy of vertex extracted by
$E_{\rm vertex}=E_\infty-E_0$.}
\label{junc.tbl}
\begin{center}
\begin{tabular}{c|rrr|rr|r}
\hline
\hline
$(k_1:k_2:k_3)$ &
$E_{6\times20^2}$ &
$E_{6\times40^2}$ &
$E_{6\times80^2}$ &
$E_\infty$ &
$E_0$ &
$E_{\rm vertex}$ \\
\hline
$(4:4:4)$ &
$130.3545$ & $130.5356$ & $130.5794$ & $130.5939$ &
$130.5936$ & $+0.0003$ \\
\hline
$(3:4:5)$ &
$127.3866$ & $127.5686$ & $127.6127$ & $127.6273$ &
$127.6270$ & $+0.0003$ \\
\hline
$(2:5:5)$ &
$121.9670$ & $122.1716$ & $122.2217$ & $122.2383$ &
$122.2382$ & $+0.0001$ \\
\hline
\end{tabular}
\end{center}
\end{table}
We find the error is of the order of $10^{-5}E_w$,
where $E_w=\Omega_3R^3T_{{\rm D}3}=19.74$ is the energy of a wrapped brane,
which represents the order of the result
that we expect on the basis of naive considerations.

\subsection{$p=4$}\label{d4sol.sec}
The dual geometry of the four-dimensional pure Yang-Mills theory
is obtained by setting $p=4$ and $b=1$,
and the dual to a QCD junction is a branched D4-brane tube.
The results of the numerical analysis for the energies of
baryon vertices is
presented in Table \ref{s4junc.tbl}.
We use $R^4T_{{\rm D}4}$ as the unit of energy.
It is seen that the result is always negative, and the absolute value is
much smaller than
the energy of a D4-brane wrapped on ${\bf S}^4$,
$E_w=\Omega_4R^4T_{{\rm D}4}=26.32$.
Even at the maximum point with $k_1=k_2=k_3$,
it is only $3.321\%$ of $E_w$.
The charge dependence of the energy is graphically depicted in
Fig. \ref{trig45.eps} (a).
\begin{table}[htb]
\caption{
Results of numerical analysis for the baryon vertex energy
in the $p=4$ case.
(Refer to the caption of Table \ref{junc.tbl}
for the meaning of each column.)
}
\label{s4junc.tbl}
\begin{center}
\begin{tabular}{c|rrr|r|r|r}
\hline
\hline
$(k_1:k_2:k_3)$ & $E_{6\times 20^2}$ & $E_{6\times40^2}$ & $E_{6\times80^2}$ & $E_\infty$ & $E_0$ & $E_{\rm vertex}$ \\
\hline
$(4:4:4)$ & $208.634$ & $209.417$ & $209.611$ & $209.676$ & $210.552$ & $-0.874$ \\
$(3:4:5)$ & $204.338$ & $205.104$ & $205.293$ & $205.356$ & $206.165$ & $-0.809$ \\
$(3:3:6)$ & $195.734$ & $196.441$ & $196.616$ & $196.674$ & $197.392$ & $-0.718$ \\
$(2:5:5)$ & $195.738$ & $196.506$ & $196.695$ & $196.758$ & $197.392$ & $-0.634$ \\
$(2:4:6)$ & $191.432$ & $192.157$ & $192.337$ & $192.397$ & $193.006$ & $-0.609$ \\
$(2:3:7)$ & $178.501$ & $179.111$ & $179.261$ & $179.311$ & $179.846$ & $-0.535$ \\
$(2:2:8)$ & $156.907$ & $157.363$ & $157.474$ & $157.511$ & $157.914$ & $-0.403$ \\
$(1:5:6)$ & $178.484$ & $179.278$ & $179.473$ & $179.538$ & $179.846$ & $-0.308$ \\
$(1:4:7)$ & $169.861$ & $170.553$ & $170.723$ & $170.779$ & $171.073$ & $-0.294$ \\
$(1:3:8)$ & $152.576$ & $153.098$ & $153.224$ & $153.266$ & $153.527$ & $-0.261$ \\
$(1:2:9)$ & $126.591$ & $126.904$ & $126.977$ & $127.002$ & $127.208$ & $-0.206$ \\
$(1:1:10)$ & $91.830$ &  $91.961$ &  $91.989$ &  $91.998$ &  $92.116$ & $-0.118$ \\
\hline
\end{tabular}
\end{center}
\end{table}

\subsection{$p=5$}
\begin{figure}[htb]
\centerline{\includegraphics[width=.6\linewidth]{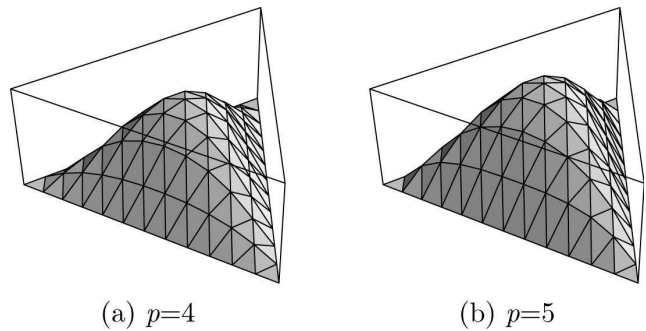}}
\caption{3D-plot of the function $-E_{\rm vertex}(k_1,k_2,k_3)$ for $p=4$ and $5$.
Various combinations of the three string charges satisfying $k_1+k_2+k_3=N$
are represented as points in the triangle.
The distance from a point to any side is proportional to the corresponding string charge.
The absolute value of $E_{\rm vertex}$ is represented by
the height of the corresponding point on the surface.
}
\label{trig45.eps}
\end{figure}

The dual geometry of the three-dimensional pure Yang-Mills theory
is obtained by setting $p=5$ and $b=1$,
and the dual to a QCD junction is a branched D5-brane tube.
The results of the numerical analysis for the energies of
baryon vertices is
presented in Table \ref{s5junc.tbl}.
We use $R^5T_{{\rm D}5}$ as the unit of energy.
Here, it is found that the result is always negative and the absolute value is
much smaller than
the energy of a D5-brane wrapped on ${\bf S}^5$,
$E_w=\Omega_5R^5T_{{\rm D}5}=31.01$.
Even at the maximum point with $k_1=k_2=k_3$,
it is only $3.325\%$ of $E_w$.
\begin{table}[htb]
\caption{
Results of numerical analysis for the baryon vertex energy
in the $p=5$ case.
(Refer to the caption of Table \ref{junc.tbl}
for the meaning of each column.)
}
\label{s5junc.tbl}
\begin{center}
\begin{tabular}{c|rrr|r|r|r}
\hline
\hline
$(k_1:k_2:k_3)$ & $E_{6\times20^2}$ & $E_{6\times40^2}$ & $E_{6\times80^2}$
& $E_\infty$ & $E_0$ & $E_{\rm vertex}$ \\
\hline
$(4:4:4)$ & $280.022$  & $281.610$ & $282.004$ & $282.135$ & $283.166$ & $-1.031$ \\
$(3:4:5)$ & $274.516$  & $276.067$ & $276.451$ & $276.579$ & $277.542$ & $-0.963$ \\
$(3:3:6)$ & $263.653$  & $265.101$ & $265.460$ & $265.579$ & $266.450$ & $-0.871$ \\
$(2:5:5)$ & $263.176$  & $264.699$ & $265.076$ & $265.201$ & $265.980$ & $-0.779$ \\
$(2:4:6)$ & $257.817$  & $259.278$ & $259.638$ & $259.758$ & $260.512$ & $-0.754$ \\
$(2:3:7)$ & $241.629$  & $242.906$ & $243.220$ & $243.324$ & $244.001$ & $-0.677$ \\
$(2:2:8)$ & $214.206$  & $215.223$ & $215.468$ & $215.549$ & $216.084$ & $-0.535$ \\
$(1:5:6)$ & $239.959$  & $241.484$ & $241.858$ & $241.198$ & $242.396$ & $-0.406$ \\
$(1:4:7)$ & $229.266$  & $230.662$ & $230.999$ & $231.110$ & $231.509$ & $-0.399$ \\
$(1:3:8)$ & $207.701$  & $208.811$ & $209.078$ & $209.166$ & $209.530$ & $-0.373$ \\
$(1:2:9)$ & $174.719$  & $175.455$ & $175.619$ & $175.672$ & $175.989$ & $-0.317$ \\
$(1:1:10)$ & $129.281$ & $129.657$ & $129.731$ & $129.754$ & $129.956$ & $-0.202$ \\
\hline
\end{tabular}
\end{center}
\end{table}
The charge dependence of the vertex energy is graphically depicted in Fig. \ref{trig45.eps} (b).

\section{Conclusions and Discussions}
In this paper we computed the energies of baryon vertices
in non-supersymmetric large $N$ $\SU(N)$ gauge theories
in $d=3$, $4$ and $5$ dimensions.
We realized the QCD junctions as D-brane configurations in
AdS black hole backgrounds.
In the $d=5$ case the problem can be analytically solved, and
it is found that the
vertex energy vanishes.
For $d=3$ and $4$,
we obtained negative energies from numerical analysis.
We also found that
the absolute value of the energy is at most only a few percent of that of a
wrapped D-brane.
Even though we have obtained results on the AdS side,
their physical meaning is still not clear.
It would be interesting to study this problem
using a more direct approach in large $N$ gauge theory.

Because there is a large class of classical solutions
appearing in AdS/CFT as the duals of confining gauge theories
for which the IR subspace possesses the $M_d\times{\bf S}^n$ structure,
we can compute the vertex energy of other theories
by simply replacing the parameters $p$ and $b$.
In this paper, we studied only the case with $b=1$.
There are also classical solutions whose IR subspaces have $b\neq1$.
For example, the IR subspace of
the Klebanov-Strassler solution\cite{kn,KS,lecture}
is obtained by setting $b=0.93266\cdots$ and $p=3$.
In Ref. \citen{junc}, it was shown that the vertex energy in the Klebanov-Strassler solution is
negative.
Because we know that the vertex energy for $b=1$ and $p=3$ vanishes,
it seems that as the parameter $b$ increases,
the vertex energy also increases.
This is consistent with the fact that in the $b\rightarrow\infty$ limit
(the tough D-brane limit)
the vertex energy is simply the energy of a wrapped D-brane
and is positive.
Another example of an IR subspace with $b\neq 1$ is
the IR subspace of the $G_2$ manifold constructed in
Ref. \citen{CGLP}.
It is obtained by setting $b=0.9249\cdots$ and $p=4$.
The confining strings in this IR subspace were studied in Ref. \citen{herzog}.
(The $b$-parameter used here is equivalent to $3A/Q$ in Ref. \citen{herzog}.)
For the reason mentioned above, again in this case we
expect a negative vertex energy.

In all the examples considered above, the vertex energy is non-negative.
It may be interesting to seek classical solutions whose IR subspaces have
large $b$, because in the gauge theories dual to such classical solutions,
the QCD strings possess an interesting property.
If the parameter $b$ is larger than the critical value $b_{\rm cr}=\sqrt{p-1}$,
the extremal condition (\ref{minimumT}) for QCD strings
has more than one solution for a certain range of $k$.
In this case, there can be two different QCD strings
of the same charge, and if $k=N/2$, the tensions of these two
QCD strings are equal,
and we can consider a string consisting of
two or more domains.

The analysis performed in this paper is only valid in the large $N$ limit.
When we use the results we obtained here for analysis of
the real hadron spectrum,
we should not regard the values in Table \ref{s4junc.tbl}
as accurate predictions,
because they may include errors of order $1/N=1/3$.
A lesson learned here is that
we should not add
the energy of a wrapped brane to the energy of a hadron,
and it would be better to simply omit the vertex contribution.

\section*{Acknowledgements}
I would like to thank M.~Bando and A.~Sugamoto for valuable discussions.
This work is supported in part by
a Grant-in-Aid for the Encouragement of Young Scientists
(\#15740140)
and
a Grant-in-Aid for Scientific Research
(\#17540238)
from the Japan Ministry of Education, Culture, Sports,
Science and Technology.

%


\begin{thebibliography}{99}
\bibitem{maldacena}
    J.~M.~Maldacena,
    Adv.Theor.Math.Phys. {\bf2} (1998) 231;
    Int.J.Theor.Phys. {\bf38} (1999) 1113.
\bibitem{GKP}
    S.~S.~Gubser, I.~R.~Klebanov, A.~M.~Polyakov,
    Phys.Lett. {\bf B428} (1998) 105.
\bibitem{holography}
    E.~Witten,
    Adv.Theor.Math.Phys. {\bf 2} (1998) 253.
\bibitem{thermal}
    E.~Witten,
    Adv.Theor.Math.Phys. {\bf 2} (1998) 505.
\bibitem{wl}
    J.~M.~Maldacena,
    Phys.Rev.Lett.{\bf 80} (1998) 4859.
\bibitem{reyyee}
    S.~-J.~Rey, J.~-T.~Yee,
    Eur.Phys.J. {\bf C22} (2001) 379.
\bibitem{tension}
    C.~P.~Herzog, I.~R.~Klebanov,
    Phys.Lett. {\bf B526} (2002) 388.
\bibitem{myers}
    R.~C.~Myers,
    JHEP {\bf 9912} (1999) 022.
\bibitem{baryon}
    E.~Witten,
    JHEP {\bf9807} (1998) 006.
\bibitem{ooguri}
  D.~J.~Gross, H.~Ooguri,
  Phys.Rev. {\bf D58} (1998) 106002.
\bibitem{HS} 
    G.~Horowitz, A.~Strominger,
    Nucl. Phys. {\bf B360} (1991) 197.
\bibitem{junc}
    Y.~Imamura,
    Prog.~Theor.~Phys. {\bf112} (2004) 1061.
\bibitem{MN1}
    J.~Maldacena, C.~N\'u\~nez,
    Int.J.Mod.Phys. {\bf A16} (2001) 822.
\bibitem{MN2}
    J.~M.~Maldacena, C.~N\'u\~nez,
    Phys.Rev.Lett. {\bf86} (2001) 588.
\bibitem{kn}
    I.~R. Klebanov, N.~A.~Nekrasov,
    Nucl.Phys. {\bf B574} (2000) 263.
\bibitem{KS}
    I. R. Klebanov, M. J. Strassler,
    JHEP {\bf0008} (2000) 052.
\bibitem{lecture}
    C.~P.~Herzog, I.~R.~Klebanov, P.~Ouyang,
    {\it``D-Branes on the Conifold and N=1 Gauge/Gravity Dualities''},
    {\tt hep-th/0205100}.
\bibitem{CGLP}
    M.~Cvetic, G.~W.~Gibbons, H.~Lu, C.~N.~Pope,
    Nucl.Phys. {\bf B606} (2001) 18.
\bibitem{herzog}
    C.~P.~Herzog,
    Phys.Rev. {\bf D66} (2002) 065009.
\end{thebibliography}
\end{document}